\documentclass[reprint,showpacs,preprintnumbers,amsmath,amssymb,superscriptaddress, aps, pra]{revtex4-1}

\eprint

\usepackage{graphicx}% Include figure files
\usepackage{dcolumn}% Align table columns on decimal point
\usepackage{tabularx}
\usepackage{bm}% bold math
\usepackage{color}
\makeatletter
\def\Ddots{\mathinner{\mkern1mu\raise\p@	
\vbox{\kern7\p@\hbox{.}}\mkern2mu
\raise4\p@\hbox{.}\mkern2mu\raise7\p@\hbox{.}\mkern1mu}}
\makeatother
\usepackage[pdfborderstyle={/S/U/W 1}]{hyperref} 
\usepackage{enumitem}% http://ctan.org/pkg/enumitem

\usepackage{lineno}
\begin{document}

\title{Attosecond-fast internal photoemission }% Force line breaks with \\

\author{Christian Heide} 
\email[E-mail: ]{christian.heide@fau.de}
\affiliation{Laser Physics, Department of Physics, Friedrich-Alexander-Universit\"at Erlangen-N\"urnberg (FAU), Staudtstrasse 1, D-91058 Erlangen, Germany}

\author{Martin Hauck}
\affiliation{Applied Physics, Department of Physics, Friedrich-Alexander-Universit\"at Erlangen-N\"urnberg (FAU), Staudtstrasse 7, D-91058 Erlangen, Germany}
\author{Takuya Higuchi}
\affiliation{Laser Physics, Department of Physics, Friedrich-Alexander-Universit\"at Erlangen-N\"urnberg (FAU), Staudtstrasse 1, D-91058 Erlangen, Germany}
\author{Jürgen Ristein}
\affiliation{Laser Physics, Department of Physics, Friedrich-Alexander-Universit\"at Erlangen-N\"urnberg (FAU), Staudtstrasse 1, D-91058 Erlangen, Germany}
\author{Lothar Ley}
\affiliation{Laser Physics, Department of Physics, Friedrich-Alexander-Universit\"at Erlangen-N\"urnberg (FAU), Staudtstrasse 1, D-91058 Erlangen, Germany}
\author{Heiko B. Weber}
\affiliation{Applied Physics, Department of Physics, Friedrich-Alexander-Universit\"at Erlangen-N\"urnberg (FAU), Staudtstrasse 7, D-91058 Erlangen, Germany}
\author{Peter Hommelhoff}
\email[E-mail: ]{peter.hommelhoff@fau.de}
\affiliation{Laser Physics, Department of Physics, Friedrich-Alexander-Universit\"at Erlangen-N\"urnberg (FAU), Staudtstrasse 1, D-91058 Erlangen, Germany}
\date{\today}% It is always \today, today,
       % but any date may be explicitly specified

\maketitle

\textbf{The photoelectric effect has a sister process relevant in optoelectronics called internal photoemission [1–3]. Here an electron is photoemitted from a metal into a semiconductor [4,5]. While the photoelectric effect takes place within less than 100 attoseconds (1 as = 10$^{-18}$ seconds) [6,7], the attosecond time scale has so far not been measured for internal photoemission. Based on the new method CHArge transfer time MEasurement via Laser pulse duration-dependent saturation fluEnce determinatiON – CHAMELEON –, we show that the atomically thin semi-metal graphene coupled to bulk silicon carbide, forming a Schottky junction, allows charge transfer times as fast as (300$\pm$200) attoseconds. These results are supported by a simple quantum mechanical model simulation. With the obtained cut-off bandwidth of 3.3 PHz (1 PHz = 10$^{15}$ Hz) for the charge transfer rate, this semimetal-semiconductor interface represents a functional solid-state interface offering the speed and design space required for future light-wave signal processing.}

The transfer of charge via internal photoemission at a solid-state interface is a fundamental process with direct relevance in ultrafast optoelectronics [2–4] and the transduction of light to chemical or electrical energy in light-harvesting [8–11]. Various solid state-based interfaces have been investigated to study the ultimate speed of this fundamental process using optical methods [2,3,12,13]. Time constants for the charge transfer in the attosecond domain ($\sim$100 as) have so far only been observed in photoemission from metal surfaces or atoms into vacuum [6,7,14] or from atoms/molecules to metals [15–17]; part of these experiments infer the charge transfer rates based on the uncertainty principle [12,18]. Similarly, excited charges may oscillate within a molecule from one part of the molecule to another within hundreds of attoseconds [10]. In stark contrast, in materials relevant for fast electronics, such as layered heterostructures with atomically sharp interfaces, the hitherto fastest charge transfer time reported was about 7 fs in a graphene–boron nitride–graphene layered heterostructures [12]. Achieving faster time scales has proven impossible so far because of the formation of excitons, which are bound states of the photo-generated electron-hole pair [2,8], or quantum mechanical backreflection [10], taking precedence over charge separation.
The ideal system to achieve attosecond-fast charge separation at a solid-state interface resembles the photoelectric effect, so external photoemission from a metal into vacuum [6,7]: It is strongly asymmetric with only one side optically absorbing; it has an atomically sharp interface, reducing the electron transfer distance; in addition to its external counterpart it has a strong built-in electric field that promotes fast transfer of electrons into the electron-absorbing half-space, i.e., the acceptor material. Electron absorption at such an extended acceptor might preserve electronic coherence but hinders the carrier wavefunction to sling back to the donor material [2,3,8,10]. 
\begin{figure}[h!]
	\begin{center}
		\includegraphics[width=8cm]{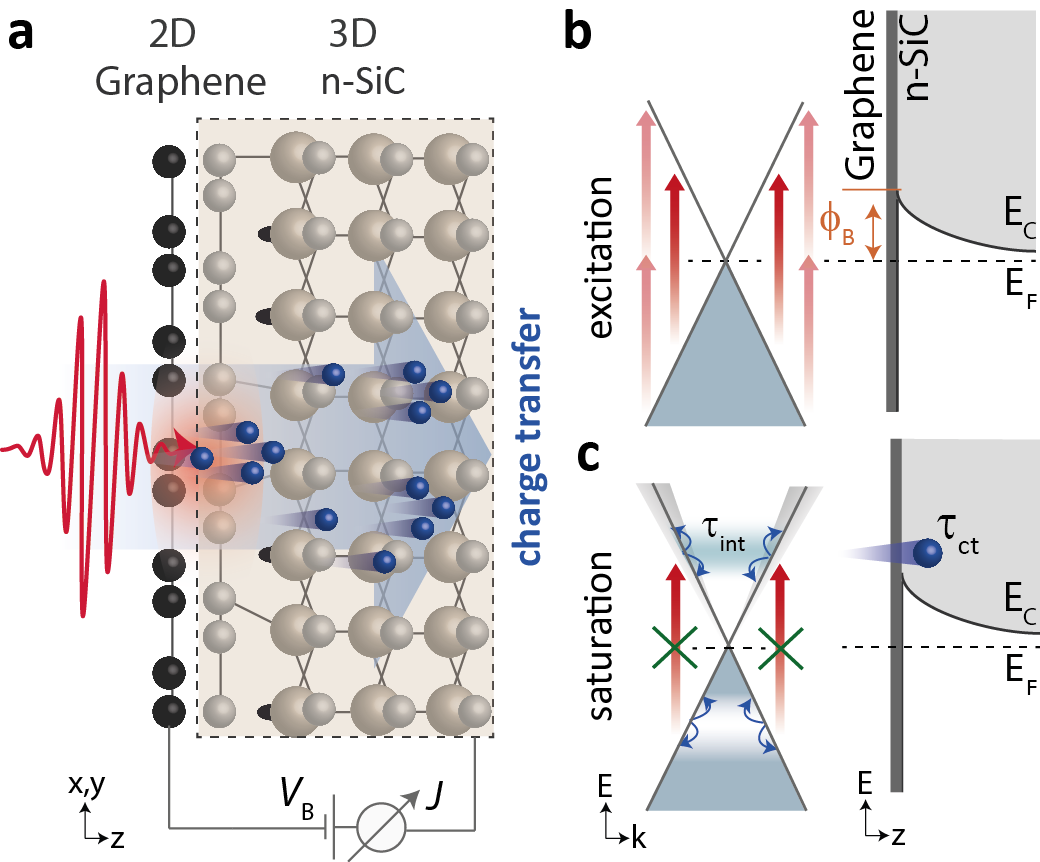}% Here is how to import EPS art
		\caption{\textbf{Experimental configuration and photocurrent generation mechanisms.} \textbf{a}, Monolayer graphene on n-doped 4H silicon carbide (n-SiC): a one-atom-thin (semi-) metallic layer coupled to a volume semiconductor. This peculiar 2D–3D material combination enables attosecond charge transfer times. We measure the photocurrent from graphene to SiC resulting from illumination with femtosecond laser pulses. \textbf{b}, \textbf{c}, Sketch of the electronic band structure of graphene in k-space representation (left) and band alignment of graphene on n-SiC in real-space representation (right). The Fermienergy is labeled as $E_\text{F}$ and the bottom of the conduction band edge in SiC as $E_\text{C}$. b, 1-photon and 2-photon absorption excite electrons in graphene. c, Excited electrons relax via two mechanisms: electrons can proceed over the Schottky barrier $\Phi_\text{B}$ into SiC within the charge transfer time $\tau_\text{CT}$, called prompt internal photoemission (PIPE or 2P-PIPE), or relax intrinsically towards a Fermi-Dirac distribution, with a time constant $\tau_\text{int}$. Optical absorption is saturated (indicated by the crossed-out excitation arrows) when the excited states are fully populated (or the initial states depopulated). When the charge transfer becomes attosecond fast, time-energy uncertainty demands broadening of the energy bands, which we indicate by the grey shading. The SiC valence band is not visible in these graphs because its energy lies below the plotting range.}
	\end{center}
	\vspace{-0.5cm}
\end{figure}

The 2D semimetal graphene grown epitaxially on the wide-bandgap semiconductor 4H silicon carbide (SiC) represents such an ideal system (Fig. 1a, [5]). With a SiC bandgap of 3.2 eV and a photon energy ranging from 1.3 eV to 2.0 eV, electrons are excited from the valence (VB) to the conduction band (CB) in graphene, when this metal-dielectric interface known as a Schottky junction, is illuminated by ultrashort laser pulses. Facing a barrier height of $\Phi_\text{B}$= 0.8 eV, most excited electrons can directly overcome the energy barrier and generate an interlayer photocurrent (Figs. 1b, 1c). In contrast to van-der-Waals heterostructures [12,19], the 2D-3D material system with a Schottky barrier used here provides a strong built-in electric field, supporting ultrafast charge transfer. We note that an interface consisting of one material (here 2D graphene) coupled to a different material (here 3D SiC) allows direct charge transport measurements, which is different from charge transfer from a metal surface coupled to bulk [20].
We introduce a surprisingly powerful method to measure the attosecond fast charge transfer time: charge transfer time measurement via laser pulse duration-dependent saturation fluence determination – CHAMELEON. CHAMELEON is based on saturable absorption in graphene and the resulting saturation of the photocurrent [21–23]. Absorption saturates when electrons are excited from the valence into the conduction band at a rate that results in the valence (conduction) band states being substantially emptied (filled). The depopulation of occupied conduction band states takes place via two channels: charge transfer from graphene to SiC with a time constant $\tau_\text{CT}$ or via intrinsic depopulation within graphene with a time constant of $\tau_\text{int}$ (Fig. 1b, 1c). Whereas the first process strongly depends on the electric field at the interface, the latter relies on electron–electron scattering with a time constant $\tau_\text{ee} \sim$ (13-80) fs [24–26], and electron-phonon and phonon-phonon scattering with a time constant of $\tau_\text{cool} \sim$ (0.7 – 3) ps [26,27]. Hence, the competition between the excitation rate and both empty VB state replenishment and population decay of CB states determines the saturation laser fluence $F_\text{s}$, at which saturation is reached. Note that $\tau_\text{CT}$ does not include the time scale associated with the electronic excitation. Thus Chameleon differs from streaking7 or attoclock measurements [15–17], which are typically pump-probe measurements including the electronic excitation. 
\begin{figure*}[t!]
	\begin{center}
		\includegraphics[width=17cm]{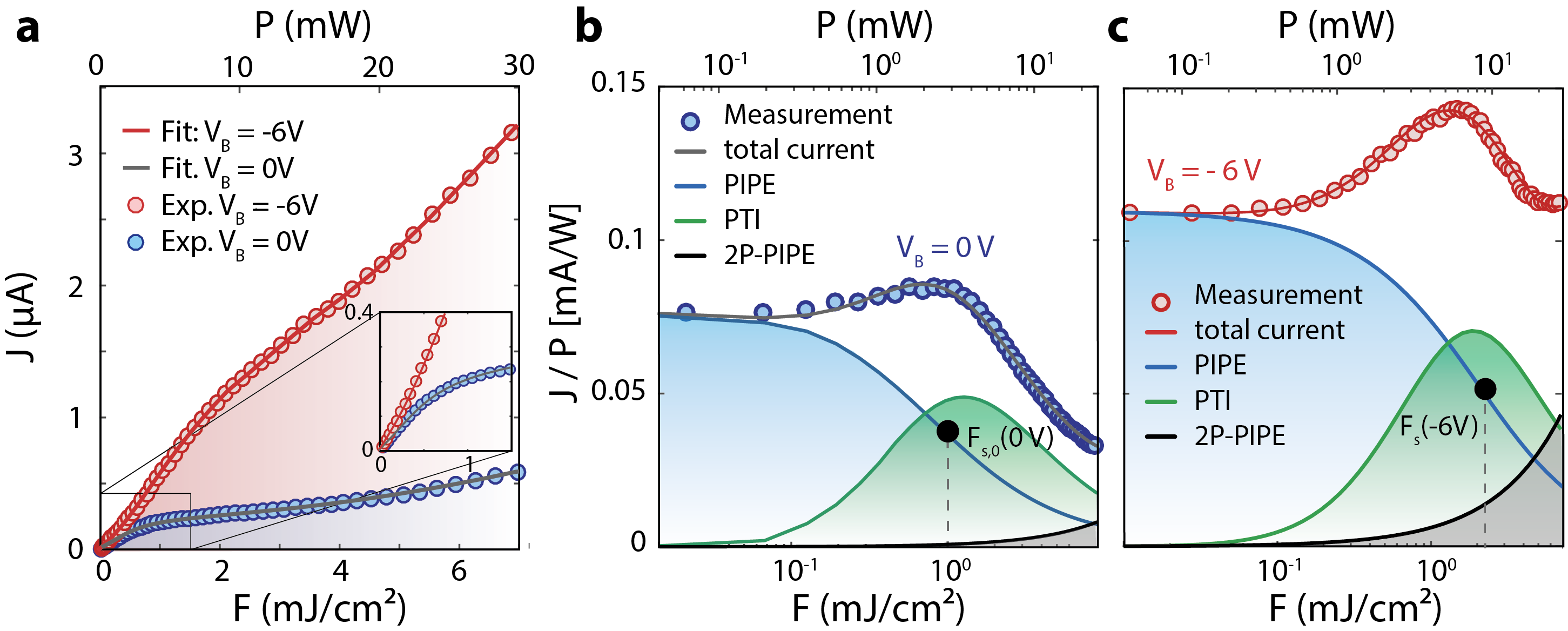}% Here is how to import EPS art
		\caption{\textbf{Measured photocurrent and model simulation results.} \textbf{a}, Photocurrent $J$ as a function of incident laser fluence $F$ (bottom axis) and average laser power $P$ (top axis) without bias voltage (blue) and with $V_\text{B}$ = -6 V (red) for $\tau_\text{P}$ = 6 fs. The inset shows the initial linear regime in both curves (up to $\sim$ 0.4 mJ/cm$^2$).  The solid lines show the model fit results, based on Eq. (13) of the SI. \textbf{b}, \textbf{c}, Efficiency ($J/P$) as a function of F and P in a semi-log plot for  $V_\text{B}$ = 0 V in b and $V_\text{B}$ = -6 V in c. The contributions of PIPE, PTI, and 2P-PIPE to the total efficiency are shown by the blue, green, and black lines, respectively. The saturation fluence (indicated by the black dot) increases from $F_\text{(s,0)}$  = 1.1 mJ/cm$^2$ without bias to $F_\text{s}$= 2.3 mJ/cm$^2$ for the biased case. An additional nonlinear contribution due to 2P-PIPE appears at higher fluence. Note that $J/P$ shows a local minimum at $\sim$ 5 mJ/cm$^2$ and rises again for larger values (not shown.)}
	\end{center}
	\vspace{-0.5cm}
\end{figure*}
In the experiment, we illuminate the graphene-SiC interface with femtosecond laser pulses with a center photon energy of 1.5 eV, a repetition rate of 80 MHz and a Fourier-limited pulse duration of 6 fs focused to a spot with 1.5 $\mu$m radius. We measure the photocurrent $J$ through the interface for various applied bias voltages $V_\text{B}$. Increasing the reverse bias voltage results in an increase of the built-in electric field and a faster charge transfer time (Extended Data Fig. 1). The excited electrons can go over or tunnel through the energy barrier within $\tau_\text{CT}$, a process called prompt internal photoemission [PIPE or 2-photon PIPE (2P–PIPE), Figs. 1b]. Alternatively, the electrons may thermalize towards a Fermi-Dirac distribution via electron-electron and electron-phonon scattering. Its high energy tail contributes a photo-thermionic current (PTI) component to $J$ [12]. 
We have developed a model based on rate equations to treat PIPE combined with a two-temperature model for graphene to cover PTI [28]. Importantly, PIPE scales linearly with laser fluence, whereas PTI scales superlinearly [12,29]. Due to the different scaling laws, we can disentangle PIPE and PTI, determine $F_\text{s}$ and attain the charge transfer time as a function of bias voltage. The model is discussed in detail in the SI.

Figure 2a shows two example photocurrent measurements as a function of F: blue data points for zero bias voltage and red points for $V_\text{B}$ = -6 V. Below $F \sim$ 0.4 mJ/cm$^2$, the photocurrent $J$ scales nearly linearly with $F$ in both cases and can, therefore, be attributed to PIPE. With increasing laser fluence, $J$ deviates from linear scaling, in subtle and different ways for the two measurements. To highlight these deviations and differences, we plot the efficiency, defined as $J$ divided by the average laser power P, in Figs. 2b and 2c on a logarithmic scale in $F$ for the two cases. The data are fitted to the sum of PIPE and PTI by employing the model outlined in SI. The two contributions to the photocurrent are also plotted individually in the figures.
The variation in the efficiency J/P with fluence is the result of both processes with different weights and different saturation fluences. For $V_\text{B}$ = 0 (Fig. 2b) the PTI contribution serves to slightly enhance J/P before it drops off, and from the fit we obtain $F_\text{s}$ = 1.1 mJ/cm$^2$ (black dot). For $V_\text{B}$=-6 V (Fig. 2c), PTI leads to a substantial maximum in efficiency around 1 mJ/cm$^2$ before it drops off at higher fluences. Here the saturation fluence rises to $F_\text{s}$ = 2.3 mJ/cm$^2$. The model fit attributes this behaviour to the superposition of PIPE and PTI contributions, which both saturate at the same laser fluence of $F_\text{s}$ = 2.3 mJ/cm$^2$. This indicates that their common origin is the saturation of the 1-photon absorption channel. For even higher laser fluences, both data sets exhibit a second superlinear, non-saturated contribution, which results from prompt internal 2-photon emission (2P-PIPE, black curve in Figs. 2b, 2c). The solid lines through the data points in Fig. 2 are the result of model simulations, which show excellent quantitative agreement with the experiment. For this reason, we trust the simulation to yield proper Fs values, which we use to analyse larger data sets (i.e. Extended Data Fig. 2).

\begin{figure*}[t!]
	\begin{center}
		\includegraphics[width=17cm]{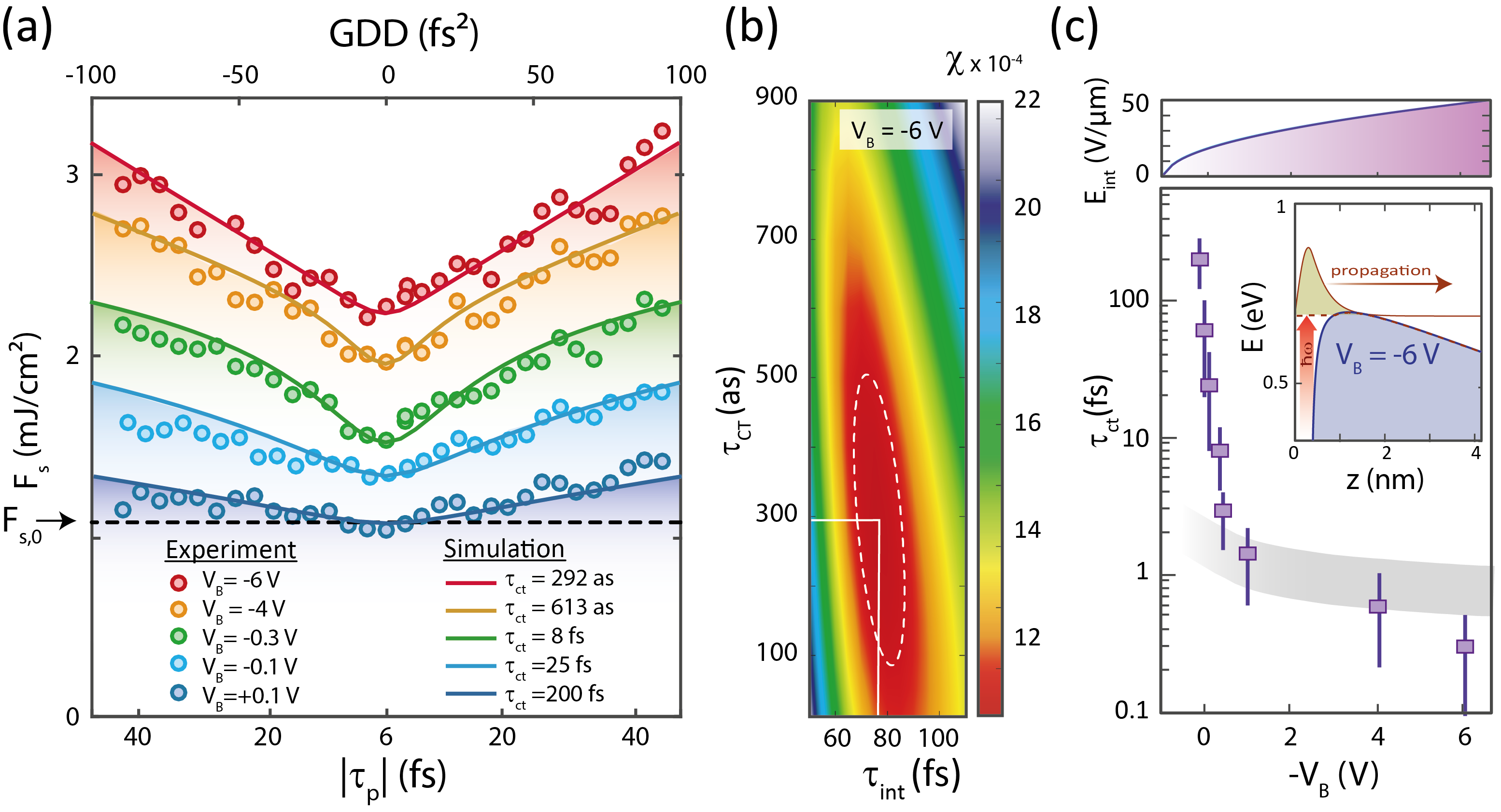}% Here is how to import EPS art
		\caption{\textbf{Extraction of the charge transfer times from graphene to SiC.} \textbf{a}, Measured (data points) and simulated (solid lines) saturation fluence $F_\text{s}$ as a function of $\tau_\text{P}$. For $V_\text{B}$ = 0.1 V (dark blue, see legend), $F_\text{s}$ is almost constant for $\tau_\text{P}$ = (6 – 45) fs. For increasing reverse bias, $F_\text{s}$ becomes more and more dependent on $\tau_\text{P}$. At the extreme case of $V_\text{B}$=-6 V, $F_\text{s}$ strongly depends on the pulse duration over the entire $\tau_\text{P}$ range, indicating a charge transfer faster than $\tau_\text{P}$. For  $V_\text{B}$=-6 V, best agreement with simulation is obtained for $\tau_\text{CT}$= (292$\pm$210) as and $\tau_\text{int}$ = (73$\pm$11) fs. Note that the pulse duration axis is symmetric around the Fourier-limited pulse duration of 6 fs because of negative and positive group delay dispersion (GDD, top axis, up-chirp and down-chirp).  b, Confidence map for $F_\text{s}$ as a function of  $\tau_\text{CT}$ and $\tau_\text{int}$ for $V_\text{B}$=-6 V. The white dotted ellipse contains the 1-sigma confidence area. $\chi$ 
		represents the goodness of fit. See Supplementary Informaiton for definition. c, Charge transfer time as a function of bias voltage $V_\text{B}$. For $V_\text{B}$= 0.1 V, the charge transfer of (210$\pm$80) fs is rather slow. With increasing reverse bias voltage, the electric field at the interface $E_\text{int}$ increases (top panel), resulting in a steep drop in the charge transfer time. Mean values and error bars for each data point are extracted from confidence maps like the one shown in (b). The gray band shows the inverse of the decay rate of an initially localized electron wave function in graphene, modelled with a semi-open quantum well. A wave packet (orange) with an energy reflecting an excitation above the barrier (purple) is illustrated in the inset for $V_\text{B}$=-6 V. 
		See SI for a detailed description of the TDSE model simulation. For large reverse bias voltage, the experimental data are well reproduced. For small bias, a space charge region is built up, which we suspect to hamper efficient charge transport into the SiC conduction band, leading to a larger $\tau_\text{CT}$ than simulated.}
	\end{center}
	\vspace{-0.5cm}
\end{figure*}
The key to extracting charge transfer times is the dependence of the saturation fluence $F_\text{s}$ on the laser pulse duration$\tau_\text{P}$. When the pulse duration is increased, charge transfer during the pulse interaction counters the saturation of absorption, resulting in a larger $F_\text{s}$. Hence, we vary$\tau_\text{P}$ from 6 fs to 42 fs and extract $F_\text{s}$ for different $V_\text{B}$. We control$\tau_\text{P}$ by varying the amount of dispersive transparent material in the beam path, the group delay dispersion (GDD). As shown in Extended Data Fig. 2, Fs shifts to larger values for increased $V_\text{B}$ and$\tau_\text{P}$ (black dots). This increase in $F_\text{s}$ is expected when the laser pulse duration exceeds the charge transfer time. 
Figure 3a shows the extracted $F_\text{s}$ as a function of VB and$\tau_\text{P}$. For $V_\text{B}$ = 0.1 V, $F_\text{s}$ hardly depends on$\tau_\text{P}$ up to 40 fs, indicating that the population decay time is longer than$\tau_\text{P}$. When the reverse bias voltage is increased, $F_\text{s}$ becomes strongly affected by the pulse duration, indicating that $\tau_\text{CT}$ is reduced to around and below$\tau_\text{P}$. From the model fit, we obtain a charge transfer time of (210$\pm$80) fs for $V_\text{B}$= 0.1 V, which steeply depends on $V_\text{B}$: for $V_\text{B}$ = -0.1 V we observe $\tau_\text{CT}$= (25$\pm$17) fs, and (8$\pm$4) fs for $V_\text{B}$= -0.3 V. Strikingly, for $V_\text{B}$ = -4 V, we measure $\tau_\text{CT}$= (613$\pm$380) as, and for the maximum applied reverse bias voltage of $V_\text{B}$= -6 V $\tau_\text{CT}$= (292$\pm$210) as.
The confidence map in $\tau_\text{CT}$ – $\tau_\text{int}$ space for $V_\text{B}$= -6 V (Fig. 3b) shows that the charge transfer occurs on a timescale of $\tau_\text{CT}$= (300$\pm$200) as, implying that this value not just represents an upper bound. An even shorter $\tau_\text{CT}$ would lead to a larger $F_\text{s}$, which would be observable with our setup. The value for the intrinsic time constant $\tau_\text{int}$ = (73$\pm$11) fs so obtained lies well within the range reported in the literature [24,25]. 

Figure 3c shows the charge transfer times as a function of bias voltage. The data displays two regimes: For large reverse bias voltages ($\sim$ 1 … 6V), the observed charge transfer time is nearly constant in the sub-femtosecond rangeand governed by electrons directly overcoming the potential barrier. For small reverse bias voltages, the charge transfer time rises steeply up to 200 fs and strongly depends on the applied voltage, which likely originates from space-charge effects hampering an efficient charge transfer.
We have also investigated the charge transfer by a simple time-dependent Schrödinger equation (TDSE) model calculation: Graphene is represented by a potential well confining the initial wavefunction such that its width matches the atomic thickness of graphene (3.2 $\AA$, see inset in Fig. 3c). The laser excitation is modelled by choosing a proper starting condition for the subsequent coherent electron dynamics, while the SiC bulk is represented by a sufficiently large continuum with the Schottky barrier potential for the conduction band electrons (see Supplementary information for details). From this model, we obtain a charge transfer time of 580 as for -6V, which matches the classical results fairly well given the simplistic nature of our model. The charge transfer follows an exponential decay rate (see SI), indicating the validity of our rate equation model. Given the quantum nature of the TDSE model, the uncertainty relation is automatically fulfilled. We note that the extremely fast charge transfer time originates from the initial strong localization of the electron wave packet to one atomic layer (graphene) coupled to a quasi-continuum (SiC). For small $V_\text{B}$, the quantum mechanical model results (gray band in Fig. 3c) clearly deviate from the single-particle TDSE approach because space charge effects (multi-electron effects) are not included in this one-electron model.
Our results, combined with the fast-growing toolbox of light-field-based electron control inside matter [1,30–33], offer new perspectives for applications and fundamental questions alike: For future petahertz electronics [1], our work represents the first passive functional element, a Schottky junction with an intrinsic bandwidth of 3.3 PHz, which could facilitate the driving of electrons across this junction using waveform controlled optical fields. The full control over the charge transfer time in this model system might help to understand quantum coherence-enhanced light-harvesting in complex bio molecules and molecular blends [9,11]. Thus the portfolio of systems with attosecond charge transfer times, which already includes photoemission into vacuum [6,7] and charge transfer within small molecules [10], is now complemented by a robust and widely designable device and a detection scheme.

\vspace{0.8cm}
\underline{References:}
\begin{enumerate}[label={[\arabic*]}]
	\setlength{\itemsep}{-5pt}%
\item	Krausz, F., Stockman, M. I. Attosecond metrology: from electron capture to future signal processing. Nat. Photon. 8, 205–213 (2014).
\item Hong, X. et al. Ultrafast charge transfer in atomically thin MoS2/WS2 heterostructures. Nat. Nanotechnol. 9, 682–686 (2014).
\item	Tan, S. et al. Coherent Electron Transfer at the Ag/Graphite Heterojunction Interface. Phys. Rev. Lett. 120, 126801 (2018).
\item	Sze, S. M., Ng, K. K. Physics of Semiconductor Devices. (Wiley, 2006).
\item	Di Bartolomeo, A. Graphene Schottky diodes: An experimental review of the rectifying graphene/semiconductor heterojunction. Phys. Rep. 606, 1–58 (2016).
\item	Cavalieri, A. L. et al. Attosecond spectroscopy in condensed matter. Nature 449, 1029–1032 (2007).
\item	Ossiander, M. et al. Absolute timing of the photoelectric effect. Nature 561, 374–377 (2018).
\item	Fassioli, F., Dinshaw, R., Arpin, P. C., Scholes, G. D. Photosynthetic light harvesting: excitons and coherence. J. R. Soc. Interface 11, 20130901 (2013).
\item 	Falke, S. M. et al. Coherent ultrafast charge transfer in an organic photovoltaic blend. Science 344, 1001–1005 (2014).
\item Wörner, H. J. et al. Charge migration and charge transfer in molecular systems. Struct. Dyn. 4, 061508 (2017).
\item	Scholes, G. D. et al. Using coherence to enhance function in chemical and biophysical systems. Nature 543, 647–656 (2017).
\item	Ma, Q. et al. Tuning ultrafast electron thermalization pathways in a van der Waals heterostructure. Nat. Phys. 12, 455–460 (2016).
\item	Zheng, Q. et al. Phonon-Assisted Ultrafast Charge Transfer at van der Waals Heterostructure Interface. Nano Lett. 17, 6435–6442 (2017).
\item	Eckle, P. et al. Attosecond Ionization and Tunneling Delay Time Measurements in Helium. Science 322, 1525–1529 (2008).
\item	Föhlisch, A. et al. Direct observation of electron dynamics in the attosecond domain. Nature 436, 373–376 (2005).
\item	Menzel, D. Ultrafast charge transfer at surfaces accessed by core electron spectroscopies. Chem. Soc. Rev. 37, 2212–2223 (2008).
\item	Borges, B. G. A. L., Roman, L. S., Rocco, M. L. M. Femtosecond and Attosecond Electron Transfer Dynamics of Semiconductors Probed by the Core-Hole Clock Spectroscopy. Top. Catal. 62, 1004-1010 (2019).
\item	Brühwiler, P. A., Karis, O., Martensson, N. Charge-transfer dynamics studied using resonant core spectroscopies. Rev. Mod. Phys. 74, 703–740 (2002).
\item	Hong, X. et al. Ultrafast charge transfer in atomically thin MoS 2 / WS 2 heterostructures. Nat. Nanotechnol. 9, 682–686 (2014).
\item	Höfer, U., Echenique, P. M. Resolubility of image-potential resonances. Surf. Sci. 643, 203–209 (2016).
\item	Marini, A., Cox, J. D., García De Abajo, F. J. Theory of graphene saturable absorption. Phys. Rev. B 95, 125408 (2017).
\item	Winzer, T. et al. Absorption saturation in optically excited graphene. Appl. Phys. Lett. 101, 221115 (2012).
\item	Xing, G., Guo, H., Zhang, X., Sum, T. C., Huan, C. H. A. The Physics of ultrafast saturable absorption in graphene. Opt. Express 18, 4564–4573 (2010).
\item	Tielrooij, K. J. et al. Generation of photovoltage in graphene on a femtosecond timescale through efficient carrier heating. Nat. Nanotechnol. 10, 437–443 (2015).
\item	Gierz, I. et al. Snapshots of non-equilibrium Dirac carrier distributions in graphene. Nat. Mater. 12, 1119–1124 (2013).
\item	Johannsen, J. C. et al. Direct view of hot carrier dynamics in graphene. Phys. Rev. Lett. 111, 027403 (2013).
\item	Lui, C. H., Mak, K. F., Shan, J., Heinz, T. F. Ultrafast photoluminescence from graphene. Phys. Rev. Lett. 105, 127404 (2010).
\item	Liang, S. J., Ang, L. K. Electron thermionic emission from graphene and a thermionic energy converter. Phys. Rev. Applied 3, 014002 (2015).
\item	Massicotte, M. et al. Photo-thermionic effect in vertical graphene heterostructures. Nat. Commun. 7, 12174 (2016).
\item	Higuchi, T., Heide, C., Ullmann, K., Weber, H. B., Hommelhoff, P. Light-field-driven currents in graphene. Nature 550, 224–228 (2017).
\item	You, Y. S., Reis, D. A., Ghimire, S. Anisotropic high-harmonic generation in bulk crystals. Nat. Phys. 13, 345–349 (2017).
\item	Langer, F. et al. Lightwave valleytronics in a monolayer of tungsten diselenide. Nature 557, 76–80 (2018).
\item	Garzón-Ramírez, A. J., Franco, I. Stark control of electrons across interfaces. Phys. Rev. B 98, 121305(R) (2018).

\end{enumerate}

\providecommand{\newblock}{}
\bibliographystyle{iopart-num}
\bibliography{MyCollection}

%\begin{figure*}
%	\begin{center}
%		\includegraphics[width=15cm]{Fig3.pdf}% Here is how to import EPS art
%		\caption{\label{Figure3}(a) to (e) Electron trajectories in reciprocal space, plotted for various delays between the driving and the control pulses. The five panels correspond to the delay values marked in Fig.$\sim$\ref{Figure2} (A to E). The initial wavenumber is $k_x=k_y=0$. Clearly, the electrons sample k-space decisively different, from a line to a spiral-out-spiral-in movement.
%			(f) to (j) CEP-dependent electronic current density $\Delta\rho = \rho_\text{C}(\Phi_\text{CEP}=\pi/2)-\rho_\text{C}(\Phi_\text{CEP}=-\pi/2)$. The red areas indicate that pulses with $\Phi_\text{CEP}=\pi/2$ generate more excitation than ones with $\Phi_\text{CEP}=-\pi/2$. For \textit{t}$_\text{delay}$ = 0 or 1.15\,fs, the red areas can be found more at $k_x > 0$. In contrast, for \textit{t}$_\text{delay}$ = 0.6\,fs at $k_x < 0$, indicating a current reversal. 
%			The dotted circles indicate energies corresponding to (multi-) photon resonances. The energy difference between conduction and valence band corresponds to $\hbar \omega$ on the innermost circle, and the subsequent rings correspond to $2\hbar \omega$ and $3\hbar \omega$. (k) - (o) Electronic current density integrated along $k_y$. The center of mass of the CEP-dependent conduction band population is indicated by the green arrow. The resulting current is plotted in Fig.$\sim$\ref{Figure2} as dashed line.}
%	\end{center}
%\end{figure*}

\end{document}